# Hardware Probing Interface and Test Robustness

Alexander Dorman, test engineer, Israel

**Abstract**. Hardware interface integrity test of an electronic product and validation of a tester's probing needles is an integral part of a test scenario. Integrity testing is based on a current voltage characteristic measurement, when a small voltage and ( or ) current stimuli are applied to the product pads including power supply circuitry pads, so that the product is not normally powered on. Needles validation should be used as a part of a self test maintenance scenario designed to predict deterioration of product probing.

**Introduction.** The set of a test fixture needle probes and connectors are used for **U**nit **U**nder **T**est (**UUT**) and a tester ports interconnect implementation. As a rule it is supposed that **UUT** probing is reliable, so that the portion of test failures the root cause of which is a probing deterioration is too small to be taken into account. Real world is more complicated. Sometimes **UUT** does not pass test procedures because of the probing deterioration. The result is the **UUT** testing failure, while detailed analysis of a failed **UUT** features finds no problem: it is **N**ot **T**roubles **F**ound (**NTF**) issue. It occurs due to the fact that the probing validity is not under the tester's monitoring and control. Mature testing relies on an assumption that the **UUT** mounted in a test fixture has reliable interconnect with the control and the sensing ports of the tester and the probing validity is under the supervision of the testing routine.
**NTF** failure risk should be minimized as much as possible because it takes an additional time to run retests and analyze the status of **UUT**: whether it really failed or not.

Therefore product pads and its relevant circuitry named here **UUT** hardware (**HW**) interface should be thoroughly tested before starting **UUT** functional testing as well as the execution of the functional test routines by the **UUT** embedded processors. **C**urrent **V**oltage **I**ntegrity **T**ests (**VCIT**) should be used to test **UUT HW** interface and to validate product probing. This method solves the **NFT** probing problem in the product testing.

This article is based on the author's unpublished papers from the period 1992 to 2004.

**Methods. VCIT** method is based on the assumption that a **UUT** not being normally powered on is a nonlinear multi port passive device. Vector of small voltage and ( or ) current stimuli are applied to the product pads as well as to its power supply circuitry pads, so that the product is not normally powered on. The **UUT** reaction vector, currents and (or) voltages, are acquired and processed to validate the following:

- **UUT** to a tester connection is valid (integrity of tester **HW** interface)
- **UUT** pads circuitry is in a valid state (integrity of **UUT HW** interface)

The results of this integrity test may be used for a tester maintenance procedure execution. For example in the beginning of the test session operator may be prompted to mount in a test fixture special dummy **UUT** instead of a regular product. This dummy **UUT** has specific pads interface circuitry which helps to validate tester probing deterioration during new test session initialization. Let's suppose for example that during test session any functional test in which **UUT** pads are involved failed. In this case the **VCIT HW** interface test should be called. If this **HW** interface test passed, the product may be considered as a failed one. If not and fixture needles have not been changed during specified time, operator should be asked to mount the dummy **UUT** in the test fixture. Failure of the **HW** integrity test with the dummy **UUT** indicates that:

- test fixture probe needles are worn and must be replaced by a new ones,

- tester maintenance self test with the dummy **UUT** must be run once more after the probe needles replacement,

- **UUT** failure must be considered as a detected **NTF** and product must be retested.

**HW** integrity test based on a **VCIT** method should be used also in a cleanup and ( or ) setup sections of a test cycle, when a **UUT** is not powered on yet normally.

**VCIT** may be implemented even with a standard equipment. E.g. is a tester where **UUT** is powered on by a power supply (**PS**) with a voltage and ( or ) current measurement options. Let's suppose that some **UUT** semiconductor circuitry pads are connected to a test fixture **IO** stimuli which are activated by a high logical levels while the **PS** output is off. The current injected by the test fixture **IO** outputs into the **UUT** is absorbed via **UUT** semiconductor circuitry by the **PS**. Related to this current voltage on the sense probes of **PS** may be measured. In assumption that the **UUT** pads are properly probed by the test fixture needles this voltage will not be close to zero. If for example **PS** HP66332A is used, **UUT** and tester have valid **HW** interfaces this voltage in some cases may be about 100…500 mV. On a contrary it is about only some millivolts when the product probing is not valid. This **VCIT** method utilizing standard **PS** and **IO** equipment may be applicable for some categories of **UUT HW** interface circuitry structures.

It should be taken into account that **UUT** behavior, when it is not in a normally powered on state, differs from the **UUT** behavior when it is powered on normally. **UUT** reaction on an applied stimulus may be slower, because **UUT** devices semiconductor structures absorb significant electrical charge.

In this case multiple measurement method when two or more levels of stimulus are sequentially in time applied to a **UUT** and relevant measurements are processed together has a better accuracy versus a single absolute measurement. In particular, it may be a differential measurements.

There is another category of a **UUT** pad circuitry structures where a **UUT** pad input controlling any **UUT** circuitry has a small influence on the **UUT** power circuitry in a not powered on state. But on the contrary **UUT** current consumption in a normally powered on state depends on the voltage applied to the **UUT** input pad. This dependence may be considered as a current voltage characteristic and utilized by the **VCIT** method.

If a signal applied to this **UUT** input pad is a specially tailored signal and an acquired **UUT** current consumption signal is correlated with an applied voltage stimulus tailored in the way be a relevant correlation reference signal, the correlation operation result may be a good indicator of **HW** probing interface integrity of **UUT** pad circuitry and corresponding test fixture probing.

Current voltage curve also may be under test. A set of a measurements should be done in the case of a shape testing. The purpose of the test is to check whether the measurement vector falls into the specified area or not. The boundaries of this area can be approximated by a hyper planes. Measurement vector comparison with the limits may be implemented as follows. It should be multiplied by a matrix whose columns are the unit vectors normal to the area boundary hyper planes. After that the product components (correlations) should be compared with the limits. These limits are the distances of a hyper planes from the origin.

Regular **VCIT** method is based on the instrumentation configuration, shown in Figure1. As a rule modern electronic devices used in **UUT** assembly have electrostatic discharge protection circuits, for example diodes VR, which are connected to a common circuit and **UUT** supply voltages potential e.g. VCC, as shown in Figure 1. If a **UUT** is not powered on by a tester's **PS** or on a contrary is biased by a small positive or negative voltage, current $I_p$ injection ( by the **Current Stimulus Source Signal** module ) results in a few hundred millivolts, sometimes more, of voltage $V_p$ on the pad of the **UUT**. This voltage value depends on the circuit structure of the semiconductor device connected to this pad: diodes VR of BGA device interface circuit, diode D as shown in Figure 1, transistors etc. This voltage $V_p$ may be simply measured by the **Voltage Signal Meter**, Figure 1. If the probe connection to the **UUT** pad or ( and ) **UUT HW** interface is not valid **Protection** module in Figure 1 signalizes about this event. In general case current stimuli may be applied to one group of pads of the **UUT** while the voltages can be measured at its other group of pads.

Furthermore detailed information may be derived from a voltage current characteristic: shape, parameters [1]. Informative parameter may be for example a characteristic deviation from the expected shape and parameters of this deviation. In this case **Stimulus Source Signal** is synchronized with a **Voltage Signal Meter** acquisition or correlation instrument, Figure 1. Synchronization is important in order to extract details of **UUT** pad circuitry voltage current characteristic. Current waves generated by a **Stimulus Source Signal** modules are applied to one and or more pads of a **UUT**. In general, acquired signal $V_p$ as a function of time is correlated with a reference signal (not shown in Figure 1) in order to measure an informative parameter. The shapes of stimulus and reference signals should be optimized in order to minimize measurement errors of an informative parameter. Extracted information may be used to recognize the damage of input/output circuitry of the devices assembled in the **UUT,** assembling errors such as an orientation of the BGA devices on the **UUT** board, not correct color of the light emission diode D etc., Figure 1.

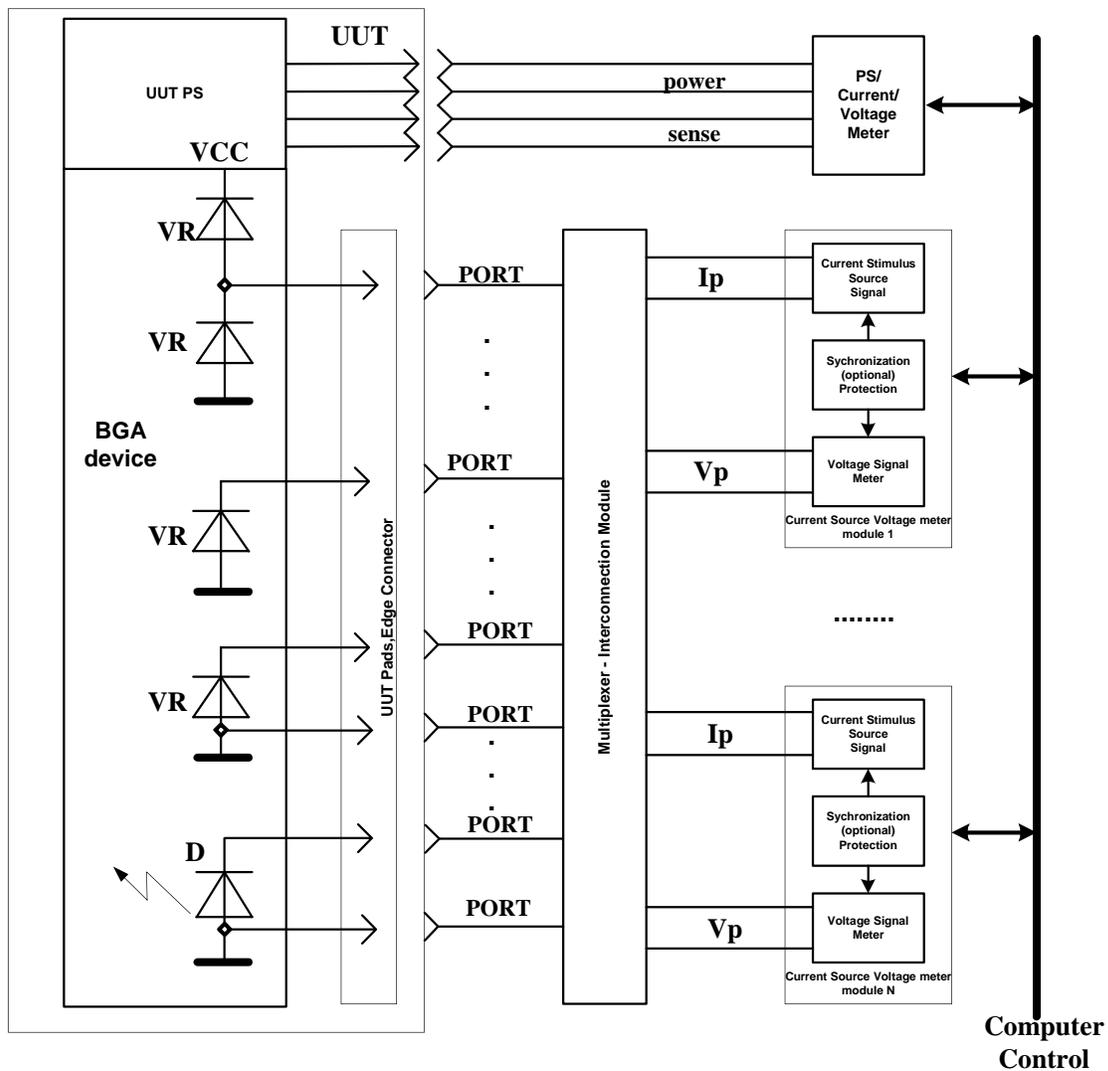

**Figure 1. Current- Voltage Integrity test instrumentation**

Aggregation of a **Current Stimulus Source Signal**, **Voltage Signal Meter**, **Synchronization and Protection** into a standard **Current Source Voltage meter** module ( **prober** ), Figure 1, may be fruitful in test fixtures design. There are some (N) **probers** in the tester, Figure 1. They are under control of the test scenario via **Computer Control** communication bus. Stimulus waveforms and reference signals, acquisition and measurement control parameters are download into a **probers** from computer of the tester.

**Conclusion. HW** probing integrity tests should be used in a test fixtures with a significant amount of the probing needles.

**VCIT** method may help to improve the test robustness and simplify the tester maintenance procedures.

Any new product introduction should be supplied with a dummy **UUT** (simulation of a correct **HW** interface integrity of real **UUT**). Dummy **UUT**, like a "golden" reference **UUT**, should be considered as an integral part of a tester.

**Prober** module may be a useful standard element of a test fixture **HW**. **Prober** module should support electrical charge - voltage/current characteristic measurements.